\documentclass{article}
\usepackage{spconf,amsmath,amssymb,graphicx}
\usepackage{gensymb}
\usepackage{glossaries}
\usepackage{booktabs}
\usepackage{multirow}
\usepackage{bm} 
\usepackage{subfloat}
\usepackage{subcaption}
\usepackage{flushend}
\usepackage{hyperref}
\usepackage[maxbibnames=4,doi=false,style=ieee,isbn=false,url=false,eprint=false]{biblatex}
\usepackage{enumitem}
\usepackage{balance}
\usepackage{graphicx}
\usepackage[font=scriptsize,labelfont=bf]{caption} 

\newcommand{\HR}{\mathrm{\mathbf{I}_{HR}}}

\newcommand{\LR}{\mathrm{\mathbf{I}_{LR}}}
\newcommand{\LRo}{\mathrm{\mathbf{I}_{LR}^{o}}}
\newcommand{\LRs}{\mathrm{\mathbf{I}_{LR}^{s}}}

\newcommand{\HRhat}{\mathrm{\mathbf{\hat{I}}_{HR}}}
\newcommand{\HRohat}{\mathrm{\mathbf{\hat{I}}_{HR}^{o}}}
\newcommand{\HRshat}{\mathrm{\mathbf{\hat{I}}_{HR}^{s}}}

\newcommand{\Net}{\mathcal{N}}
\newcommand{\Netda}{\mathcal{N}_\mathrm{\mathbf{DA}}}
\newcommand{\Netouno}{\mathcal{N}_{\mathrm{\mathbf{O}}_1}}
\newcommand{\Nets}{\mathcal{N}_\mathrm{\mathbf{S}}}
\newcommand{\Netst}{\mathcal{N}_{\mathrm{\mathbf{S}}_T}}
\newcommand{\T}{\mathcal{T}}
\newcommand{\Tda}{\mathcal{T}_\mathrm{\mathbf{DA}}}
\newcommand{\Tsinv}{\mathcal{T}_{\mathrm{\mathbf{S}}}^{-1}}
\newcommand{\Ts}{\mathcal{T}_\mathrm{\mathbf{S}}}
\newcommand{\Touno}{\mathcal{T}_{\mathrm{\mathbf{O}}_1}}

\newcommand{\Tinvda}{\mathcal{T}^{-1}_\mathrm{\mathbf{DA}}}

\newcommand{\DO}{\mathcal{D}_{\mathrm{\mathbf{O}}}}
\newcommand{\DOuno}{\mathcal{D}_{\mathrm{\mathbf{O}}_1}}
\newcommand{\DOdue}{\mathcal{D}_{\mathrm{\mathbf{O}}_2}}
\newcommand{\DS}{\mathcal{D}_{\mathrm{\mathbf{S}}}}
\newcommand{\DSt}{\mathcal{D}_{\mathrm{\mathbf{S}}_T}}
\newcommand{\simulated}{\mathrm{\mathbf{S}}}
\newcommand{\observed}{\mathrm{\mathbf{O}}}

\newacronym{voc}{VOC}{Volatile Organic Compound}
\newacronym{bvoc}{BVOC}{Biogenic Volatile Organic Compound}
\newacronym{cnn}{CNN}{Convolutional Neural Network}
\newacronym{nn}{NN}{Neural Network}
\newacronym{sr}{SR}{Super-Resolution}
\newacronym{sisr}{SISR}{Single-Image Super-Resolution}
\newacronym{san}{SAN}{Second-order Attention Network}
\newacronym{nlrg}{NLRG}{Non-Locally Enhanced Residual Group}
\newacronym{lsrag}{LSRAG}{Local-Source Residual Attention Groups}
\newacronym{srgan}{SRGAN}{Super-Resolution Generative Adversarial Network}
\newacronym{esrgan}{ESRGAN}{Enhanced SRGAN}
\newacronym{srresnet}{SRResNet}{Super-Resolution ResNet}
\newacronym{msrresnet}{MSRResNet}{Modified SRResNet}
\newacronym{rcan}{RCAN}{Residual Channel Attention Network}
\newacronym{rrdbnet}{RRDBNet}{Residual-in-Residual Dense Block Network}
\newacronym{sansisr}{SANSISR}{Second-order Channel Attention Network}
\newacronym{srcnn}{SRCNN}{SR Convolutional Neural Network}
\newacronym{cdf}{CDF}{Cumulative Distribution Function}
\newacronym{pdf}{PDF}{Probability Density Function}
\newacronym{hr}{HR}{High Resolution}
\newacronym{lr}{LR}{Low Resolution}
\newacronym{megan}{MEGAN}{Model of Emissions of Gases and Aerosols from Nature}
\newacronym{mse}{MSE}{Mean Squared Error}
\newacronym{nmse}{NMSE}{Normalized Mean Squared Error}
\newacronym{ssim}{SSIM}{Structural Similarity Index Measure}
\newacronym{dl}{DL}{Deep Learning}
\newacronym{lst}{LST}{Land Surface Temperature}
\newacronym{sst}{SST}{Sea Surface Temperature}
\newacronym{vgg}{VGG}{Visual Geometry Group}
\newacronym{pinn}{PINN}{Physics-Informed Neural Networks}
\newacronym{ctm}{CTM}{Chemistry Transport Model}
\newacronym{omi}{OMI}{Ozone Monitoring Instrument}
\newacronym{gome}{GOME-2}{Global Ozone Monitoring Experiment-2}
\newacronym{images}{IMAGESv2}{Intermediate Model of the Global and
Annual Evolution of Species}
\newacronym{form}{HCHO}{formaldehyde}
\newacronym{dam}{DAM}{Domain Adaptation Module}
\newacronym{da}{DA}{Domain Adaptation}
\newacronym{cris}{CrIS}{Cross-track Infrared Sounder}

\title{Super-Resolution of BVOC Emission Maps Via Domain Adaptation}
\name{Antonio Giganti, Sara Mandelli, Paolo Bestagini, Marco Marcon, Stefano Tubaro
\thanks{This work was supported by the Italian Ministry of University and
Research (MUR) and the European Union (EU) under the PON/REACT project.}}
\address{Dipartimento di Elettronica, Informazione e Bioingegneria - Politecnico di Milano - Milan, Italy}

\begin{document}
\ninept
\maketitle
\begin{abstract}
Enhancing the resolution of \gls{bvoc} emission maps is a critical task in remote sensing. 
Recently, some \gls{sr} methods based on \gls{dl} have been proposed, leveraging data from numerical simulations for their training process.
However, when dealing with data derived from satellite observations, the reconstruction is particularly challenging due to the scarcity of measurements to train \gls{sr} algorithms with.
In our work, we aim at super-resolving low resolution emission maps derived from satellite observations by leveraging the information of emission maps obtained through numerical simulations. 
To do this, we combine a \gls{sr} method based on \gls{dl} with \gls{da} techniques, harmonizing the different aggregation strategies and spatial information used in simulated and observed domains to ensure compatibility. 
We investigate the effectiveness of \gls{da} strategies at different stages by systematically varying the number of simulated and observed emissions used, exploring the implications of data scarcity on the adaptation strategies.
To the best of our knowledge, there are no prior investigations of \gls{da} in satellite-derived \gls{bvoc} maps enhancement. Our work represents a first step toward the development of robust strategies for the reconstruction of observed \gls{bvoc} emissions.

\end{abstract}
\begin{keywords}
Biogenic Emission, BVOC, Isoprene, Image Super-Resolution, Domain Adaptation
\end{keywords}

\glsresetall


\section{Introduction}
\label{sec:intro}


\glspl{bvoc} have been largely studied in the last two decades \cite{ciccioli_bvoc_2023, bvoc_exchange_2021, opacka_isoprene_2021, fu_direct_2019} 
as they play critical roles in atmospheric chemistry~\cite{cai_bvoc_scientometric_2021, guenther_model_2012}, promoting the formation of low-level ozone and secondary organic aerosols that strongly impact air quality and the planet's radiative budget~\cite{calfapietra_role_2013}.
Studying \gls{bvoc} emissions proves paramount for numerical evaluations of past, current, and future air quality and climate conditions~\cite{guenther_model_2012}.
To perform these \gls{bvoc}-based studies on climate, quantitative estimations of \gls{bvoc} emissions are required for numerical evaluations.

However, 
\gls{bvoc} measurements are often limited in space and time, and acquiring fine-grained \gls{bvoc} emission maps over a large region is costly and time-consuming~\cite{hewitt_bvoc_quantification_2011, fu_direct_2019}.
Therefore, the available \gls{bvoc} emission maps might be not enough for reliable simulations of atmospheric, climate, and forecasting models~\cite{bauwens_past_future_bvoc_mohycan_2018}.
To solve this issue, image \gls{sr} techniques can be applied to \gls{bvoc} acquisitions, in order to generate a denser spatial grid of emission maps by starting from a coarser one~\cite{giganti2023enhancing}. 

Recently, some methods have been proposed to super resolve \gls{bvoc} emission by leveraging \gls{dl} models~\cite{giganti2023enhancing, giganti2023mbsr}.
Despite their promising performance, the main limitation of the currently available approaches is that they require to be trained and tested on datasets with similar characteristics, otherwise there is strong risk of encountering data domain mismatch, resulting in unreliable \gls{bvoc} reconstructions~\cite{giganti2023enhancing}.

For instance, in many realistic scenarios, there might be the need to super-resolve \gls{bvoc} maps derived from actual satellite measurements, at the same time disposing of few data of this kind at training stage, due to the difficulty in acquiring these emissions.
A possible solution could be to train the \gls{sr} models on numerically simulated \gls{bvoc} emission data and use them to super resolve the real measurements. However, the two data domains might report different characteristics in terms of dynamic range and temporal aggregation. 

\begin{figure}[t]
\centering
\includegraphics[width=.9\columnwidth]{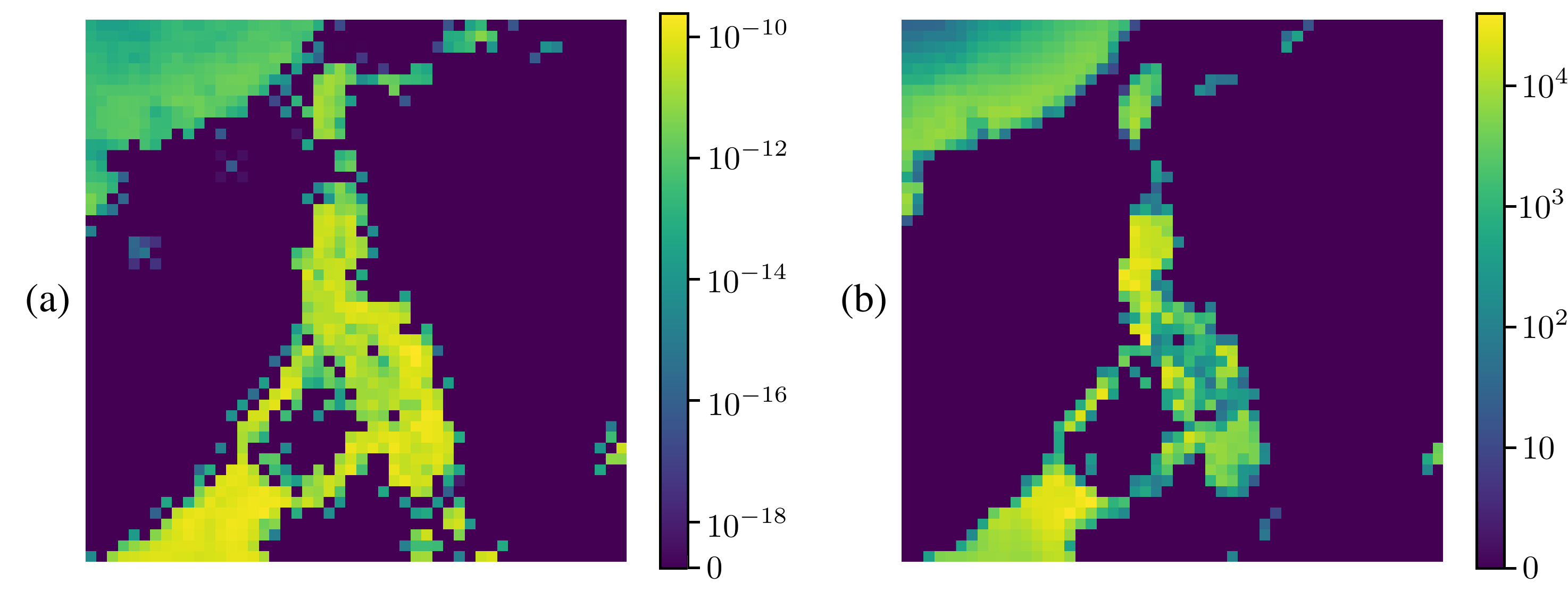}
\vspace{-10pt}
\caption{Comparison between \gls{bvoc} 
emission maps corresponding to the same geographical area but to different domains: (a) simulated data (b) observed data. Emission values and patterns differ according to the data aggregation strategies adopted~\cite{sindelarova_high-resolution_2022, de_smedt_gome_dataset_2012}.}
\label{fig:emission diff}
\vspace{-18pt}
\end{figure}

Numerically simulated data are typically aggregated over brief periods to capture short-term variability, smoothing out the long-term characteristics~\cite{sindelarova_high-resolution_2022, bauwens_past_future_bvoc_mohycan_2018}. Contrarily, satellite observations are typically aggregated over long time ranges to reduce the effects of random errors in the measurements and increase the signal-to-noise ratio~\cite{wang_long-term-bvoc_china_2021, de_smedt_gome_dataset_2012, bauwens_omi_dataset_2016}. 
In addition, climate satellite data often comes from non-geostationary satellites, binding the temporal resolution of measurements to the satellite's revisit time.
Moreover, weather conditions may be an obstacle to measurements in specific geographic areas, 
hindering the possibility
to acquire information in a short-term interval.
All these peculiar characteristics inevitably lead to a domain-shift between simulated and observed data.
For example, Fig.~\ref{fig:emission diff} depicts two \gls{bvoc} emission maps related to the same geographical area but to different data domains, i.e., simulated or observed. 
It is noticeable a smoother behaviour of observed data with respect to simulated ones which present more high spatial frequency content. Notice also the differences in the dynamic ranges.



For these reasons, we propose to investigate the capabilities of existing \gls{sr} algorithms to super resolve emission maps derived from satellite observations.  
We train the \gls{dl} models by leveraging numerically simulated data. To counteract the potential domains mismatch, we investigate the use of \gls{da}~\cite{ben-david_da_theory_2010} strategies,
with the scope of mitigating the problems that arise in dealing with data coming from a domain other than the one used in training.

Several approaches have been proposed to solve the adaptation to different domains~\cite{yadav_deepaq_2023, xu_domain_adaptation_remote_sensing_2022, peng_da_remote_sensing_survey_2022}.
However, to the best of our knowledge, no prior studies tackled \gls{da} for \gls{bvoc} emission. 
We consider the realistic condition of scarcity of observed data, pushing \gls{da} to the limits in case an extremely reduced set of satellite observations can be used to perform the \gls{da}.


\section{Domain Adaptation for Satellite-derived BVOC Enhancement}
\label{sec:proposed method}
\glsreset{hr}
\glsreset{lr}
\glsreset{sr}
\glsreset{da}

\begin{figure}[t]
\centering
\includegraphics[width=0.9\columnwidth]{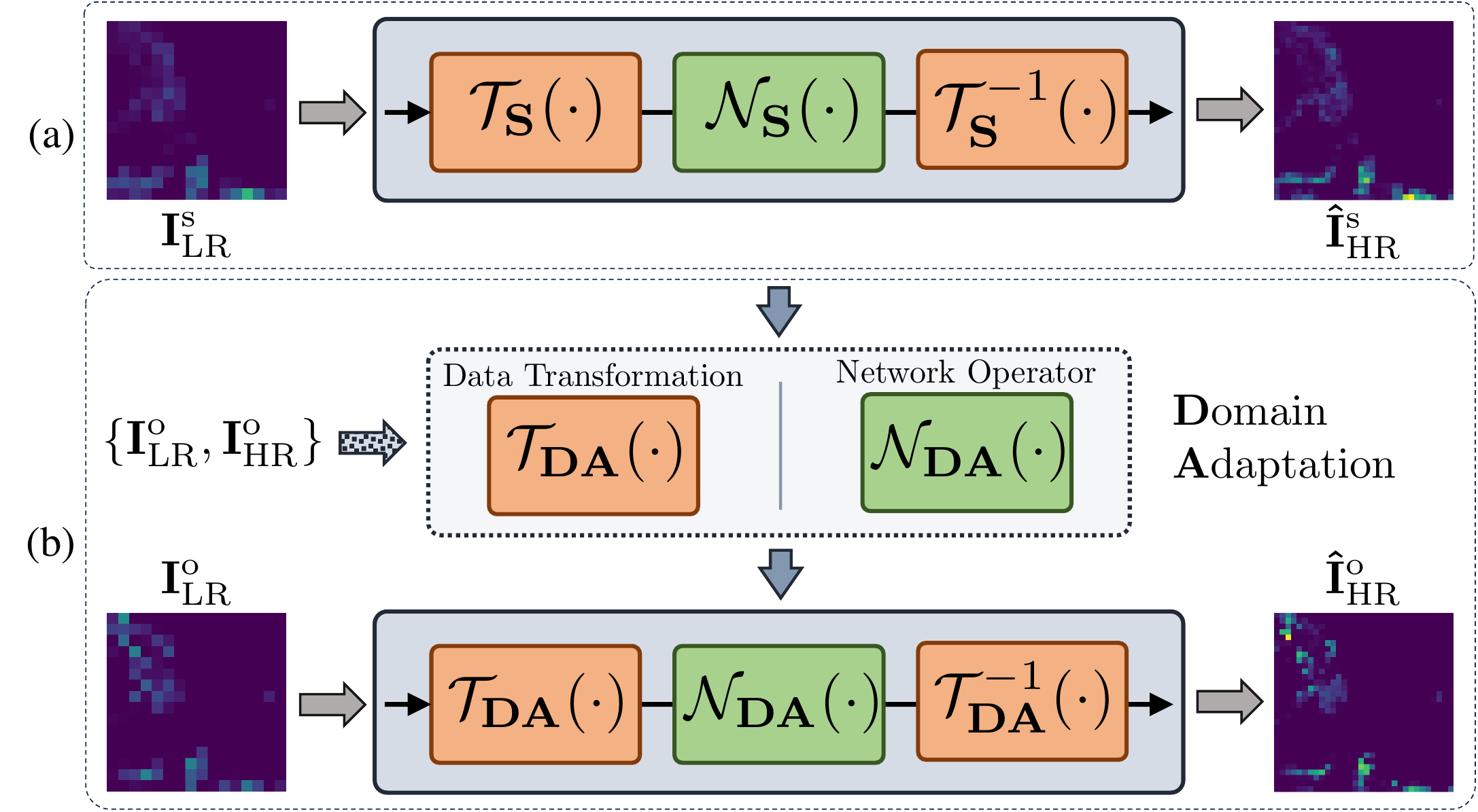}     
\vspace{-5pt}
\caption{(a) The proposed \gls{sr} system \cite{giganti2023enhancing}, which proved valid for dealing with \gls{bvoc} emission maps of $\simulated$ domain; (b) The proposed \glsfirst{da} strategies: \textit{Data Transformation} adaptation - $\Tda(\cdot)$ and \textit{Network Operator} adaptation - $\Netda(\cdot)$, that take into account the existing domain-shift between emission maps of the $\simulated$ and $\observed$ domains.}   
\label{fig:system}
\vspace{-15pt}
\end{figure}

Our goal is to super-resolve \gls{lr} \gls{bvoc} emission maps derived from satellite observations. We do so by exploiting the knowledge of emission maps obtained from numerical simulations. 
We denote with $\simulated$ the $\simulated$(\textit{imulated}) and with $\observed$ the $\observed$(\textit{bserved}) data domain. The former is obtained through numerical simulations, while the latter is derived from actual satellite observations.


In our previous work~\cite{giganti2023enhancing}, we tackled the problem of recovering \gls{hr} emission maps starting from \gls{lr} versions, considering only \textit{numerically simulated} data. We estimated \gls{hr} simulated emissions $\HRshat$ by starting from \gls{lr} emissions $\LRs$ as (see Fig.\ref{fig:system}(a)):
\begin{equation}
    \HRshat = \Tsinv(\Nets(\Ts(\LRs))),
    \label{eq:sr_simulated}
\end{equation}
being 
$\Ts$ 
a data transformation applied to \gls{lr} simulated emissions \cite{scikit-learn, peterson_oqn_2020}, 
$\Nets$
a \gls{nn} operator~\cite{dai_sansisr_2019} 
and 
$\Tsinv$
the inverse transformation of 
$\Ts$.
Both $\Ts$ and $\Nets$ operators were learnt from a training set of \textit{simulated} emissions, meaning they were tailored to the domain of data under investigation, i.e., the domain $\simulated$.



In this work, we propose to super-resolve emission maps derived from satellite \textit{observations} (domain $\observed$) by leveraging data obtained through numerical \textit{simulations} (domain $\simulated$). 
To make the most out of the information derived from simulations, 
an adaptation between the two domains becomes necessary, as we show in our experimental analysis (Sec.~\ref{sec:results}).
Fig.~\ref{fig:system}(b) shows a simplified scheme of our method.
We propose two different \gls{da} strategies: (i) data transformation adaptation; (ii) network operator adaptation. 

\noindent \textbf{Data Transformation Adaptation. }
In the first scenario, we estimate a non-parametric data transformation $\Tda$ from a pool of emissions from domain $\observed$ to adapt the different aggregation strategies and dynamic ranges of both domains and, at the same time, consider the lack of data from the $\observed$ domain compared to the $\simulated$. 
To highlight the potential advantages of our proposal, we do not modify the \gls{nn} operator, exploiting the same $\Nets$ as in \eqref{eq:sr_simulated}, i.e., the \gls{nn} trained on simulated data.
\gls{hr} observed emissions $\HRohat$ can be estimated as
\begin{equation}
\label{eq:input adaptation}
    \HRohat = \Tinvda(\Nets(\Tda(\LRo))).
\end{equation}

\noindent \textbf{Network Operator Adaptation. }
In the second scenario, we extend the \gls{da} to the \gls{nn} operator as well.
In detail, we investigate the influence of fine-tuning the \gls{nn} pre-trained on simulated data by injecting some knowledge of the target domain $\observed$. 
We estimate \gls{hr} observed emissions $\HRohat$ as
\begin{equation}
\label{eq:network adaptation}
    \HRohat = \Tinvda(\Netda(\Tda(\LRo))).
\end{equation}

In Sec.~\ref{sec:results}, we provide more details on the pursued experimental campaign, analyzing the two proposed \gls{da} strategies.


\vspace{-5pt}
\section{Dataset}
\label{sec:setup}
To validate our approach, we use different datasets belonging to the $\simulated$(\textit{imulated}) and the $\observed$(\textit{bserved}) domains. 

\noindent $\simulated$(\textbf{\textit{imulated}) Domain. } 
We use the most recent high-resolution global coverage biogenic emission inventory presented in~\cite{sindelarova_high-resolution_2022}. 
This can be considered a bottom-up approach since these data are obtained through a simulation of the \gls{megan}~\cite{guenther_model_2012}, a well-known semi-empirical modeling framework 
embedded into several Earth system and chemical transport models~\cite{cai_bvoc_scientometric_2021}.
The inventory includes emissions from \glspl{bvoc} covering the entire Earth surface in the period of 2000-2019, with a $0.25^{\circ}$ spatial resolution. Emissions are reported as hour profiles and are averaged monthly. We refer to this dataset as $\DS$ since it is obtained from numerical simulations. 

\noindent $\observed$(\textbf{\textit{bserved}) Domain. } 
We investigate the emission maps derived from two satellite instruments: (i) the \gls{gome} on the EUMETSAT-Metop-A satellite, for the period 2007-2012~\cite{de_smedt_gome_dataset_2012}; (ii) the \gls{omi} on the NASA-Aura (EOS/Chem-1) satellite, for the period 2005-2014~\cite{bauwens_omi_dataset_2016}. 
Both the top-down datasets are available on $0.50^{\circ}$ spatial resolution. Emissions are reported as daily averaged profiles.
To make a meaningful performance comparison between the two datasets, we consider emissions from the same acquisition years, thus the ones from 2007 to 2012.
We refer to the datasets from the \gls{gome} and the \gls{omi} instruments as $\DOuno$ and $\DOdue$, respectively.

\noindent \textbf{Experimental dataset. }
From all the considered datasets, we select only isoprene emission maps, being it the most contributing \gls{bvoc} in terms of global emission and atmospheric impact~\cite{opacka_isoprene_2021, sindelarova_high-resolution_2022}.

Emission maps of $\DOuno$ and $\DOdue$ datasets present a grid of $720\times360$ cells. 
The emission maps of $\DS$ instead cover the same geographical areas of the observed data but comprise $1440\times720$ cells, since the spatial resolution is twice. 
For our experiments, we computed also a version of $\DS$ with a spatial resolution of $0.50^{\circ}$, in order to match the resolution of the observed data.

We also work with a subset of $\DS$, considering emission maps of the same acquisition years of $\DO$ datasets, thus 2007-2012. 
This enables to perform more realistic comparisons with satellite-derived data. 
We refer to this subset as $\DSt$ since it is limited in \textit{Time.}


For each dataset, we slice the emission maps to obtain smaller non-overlapped patches of $32\times32$ cells, following the steps presented in our previous work~\cite{giganti2023enhancing}. 
These can be considered our ground truth \gls{hr} emission maps $\HR$.
The total number of \gls{hr} emission patches in $\DS$ is almost $630$K at $0.25^{\circ}$ and $73$K at $0.50^{\circ}$; $\DSt$ includes around $187$K \gls{hr} patches at $0.25^{\circ}$ and $22$K at $0.50^{\circ}$; each $\DO$ datasets contains almost $16$K emission patches at $0.50^{\circ}$.

We generate \gls{lr} patches by performing bicubic downsampling, obtaining emission maps $\LR$ of $16\times16$ cells. 
Our goal is estimating \gls{hr} patches from \gls{lr} ones, with an upscale factor equal to $2$. 





\vspace{-5pt}
\section{Experimental Results}
\label{sec:results}

\begin{table}[t]
\centering
\caption{Results in perfect knowledge scenario (no \gls{da} required). Training times are reported as a fraction of the time spent for the case $\DS$, $0.50^{\circ}\rightarrow 0.25^{\circ}$ (first row). }
\vspace{-5pt}
\resizebox{1\columnwidth}{!}{%
\begin{tabular}{@{}lccccc@{}}
\toprule
Dataset  &Spatial Resolution  &No.Patches &NMSE {[}dB{]} &SSIM &Training Time {[}\%{]} \\ 
\midrule
\multirow{2}{*}{$\DS$} & $0.50^{\circ}\rightarrow 0.25^{\circ}$ & $630192$  & $-23.83$ & $0.988$ & $100$ [ref]  \\
 & $1.00^{\circ}\rightarrow 0.50^{\circ}$ & $73152$ & $-19.40$ & $0.988$ & $17.60$  \\ 
 \midrule
\multirow{2}{*}{$\DSt$} & $0.50^{\circ}\rightarrow 0.25^{\circ}$ & $187488$ & $-26.54$ & $0.989$ & $23.15$  \\
 & $1.00^{\circ}\rightarrow 0.50^{\circ}$ & $21864$  & $-22.33$ & $0.989$ & $10.32$  \\ 
 \midrule
$\DOuno$ & $1.00^{\circ}\rightarrow 0.50^{\circ}$ & $16495$ & $-15.68$ & $0.979$ & $3.73$  \\ 
\midrule
$\DOdue$ & $1.00^{\circ}\rightarrow 0.50^{\circ}$ & $16495$ & $-18.07$ & $0.977$ & $3.73$ \\
 \bottomrule
\end{tabular}%
}
\label{tab:no_da_results}
\vspace{-15pt}
\end{table}

\noindent \textbf{Experimental Setup. }
When working with dataset $\DS$, we train the network $\Nets$ by adopting a $70/20/10 \, \%$ splitting strategy for train/validation/test. 
For the remaining datasets ($\DSt$, $\DOuno$ and $\DOdue$), we split the emission patches considering 2007-2010 for training, 2011 for validation, and 2012 for testing. 
We always keep an equal amount of patches in all the tests, limiting this according to the dataset with the lowest cardinality, so the $\DO$ datasets.

We evaluate the \gls{sr} performances in terms of \gls{ssim} and \gls{nmse}, defined as the \gls{mse} between $\HR$ and $\HRhat$, normalized by the average of $\HR^2$.  
A good result is the one with high \gls{ssim} and/or low \gls{nmse}. 

We implement and train our method using an NVIDIA Tesla V100 PCIe 16 GB GPU, running on an Intel Xeon Gold 6246 CPU.

\noindent \textbf{Perfect Knowledge Scenario. }
In this scenario, we investigate the performance in case a subset of \gls{hr} data of the same domain of the testing ones is available for training, i.e., no \gls{da} is needed. We use these results to show the effectiveness of the proposed \gls{sr} method on the different data domains under investigation.

To this purpose, we apply  \eqref{eq:sr_simulated} to the various datasets we experiment on, training a \gls{nn} operator $\Net$ and estimating the data transformation $\T$ from the specific training set involved. For instance, when working with $\DOuno$, we exploit its \gls{hr} training set to estimate specific $\Touno$ and $\Netouno$ operators by following our past methodology~\cite{giganti2023enhancing}.
Results achieved across the different datasets are shown in Tab.~\ref{tab:no_da_results}. 
For clarity's sake, Tab.~\ref{tab:no_da_results} includes additional information: for each dataset, we report the total number of patches in the set and the time required for training the \gls{sr} system.

For $\simulated$ datasets, i.e., $\DS$ and $\DSt$, 
the higher the spatial resolution,
the better the performance.
We believe this depends on the emissions' spatial resolution which impacts their high-frequency information. Indeed, patches at $0.50^{\circ}$ inherently contain more spatial information than those at $1.00^{\circ}$. This results in a more spatially detailed emission map.

We can notice a significant reconstruction improvement
when considering $\DSt$ with respect to $\DS$. 
Indeed, as we have shown in our previous work~\cite{giganti2023enhancing}, the splitting strategy of $\DSt$ improves the performance since the system can capture these temporal dynamics of the emission and generalize its predictions to unseen years, reflecting real-world applications.


When changing the data domain, results report a performance drop.
Focusing on the case $1.00^{\circ} \rightarrow 0.50^{\circ}$ spatial resolution, results on $\observed$ datasets are at least $4$ dB worse than the best results of $\simulated$ datasets.
We believe the emissions belonging to the $\simulated$ domain inherently contain less noise than those of $\observed$ since they are numerically simulated and, therefore, easier to describe and reconstruct.

We also evaluate the impact of the computational time required for training on the different datasets. 
On $\simulated$ domain, training on time-limited data, i.e., $\DSt$, not only leads to superior performance but also strongly reduces the computation time since a smaller number of patches is considered for training.
On $\observed$ domain, the training time is extremely reduced, due to the low cardinality of the datasets.

\noindent \textbf{Zero Knowledge Scenario. }
This scenario assumes we do not have any \gls{hr} observed data at disposal. Hence, we estimate \gls{hr} emission maps of $\observed$ domain by leveraging only the operators learnt from simulated data ($\simulated$ domain).

Our experimental analysis shows that training either on $\DS$ or $\DSt$ and testing on $\DO$ datasets results in drastic performance drops.
Indeed, the \gls{nmse} is fixed at $0$ dB for both $\DOuno$ and $\DOdue$. We believe the huge differences in data statistics from these two domains strongly impact the reconstruction of the super resolved emissions. This motivates the investigation of \gls{da} strategies.

\begin{figure}[t]
\centering
\includegraphics[width=1\columnwidth]{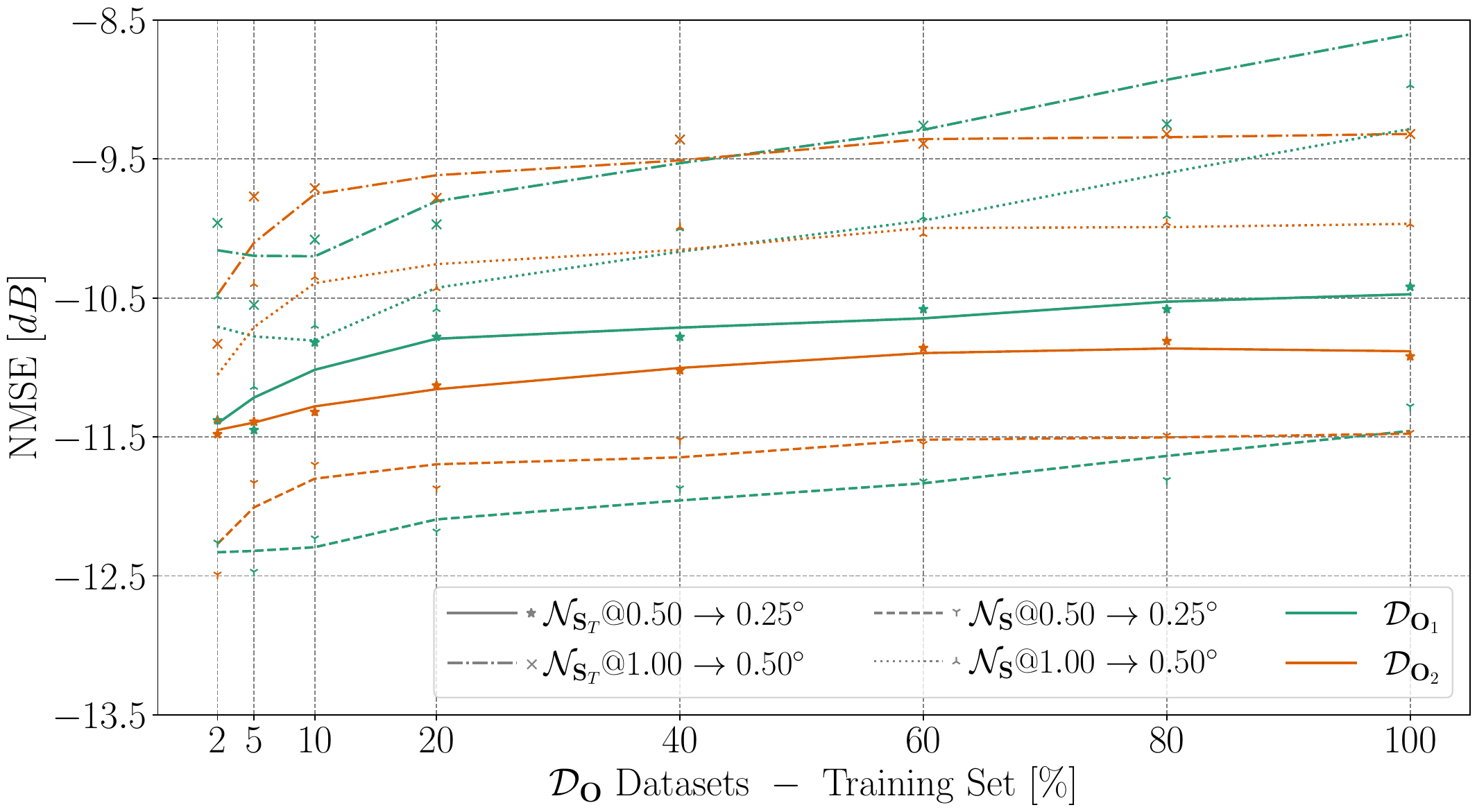}
\vspace{-17pt}
\caption{Average \gls{nmse} achieved 
by varying the amount of patches of the $\observed$ domain used to train the $\Tda$ emission transformation. The point values represent the mean value, averaged over three random subsets of the $\DO$ datasets, for each percentage. Lines report the smoothed trends of these experiment.}
\label{fig:data adaptation}
\vspace{-15pt}
\end{figure}

\noindent \textbf{Data Transformation Adaptation. }
This strategy
supposes that a certain portion of observed data is available to the analyst.
These data can be exploited to estimate a non-parametric transformation $\Tda$ characterizing the emission statistics of the $\observed$ domain.
In specific, we investigate how many data from this domain impact the adaptation process, 
considering an increasing percentage of the $\DOuno$ and $\DOdue$ training partitions for estimating the $\Tda$ transformation. 

As reported in Section~\ref{sec:proposed method}, we keep the \gls{nn} operator trained on the $\simulated$ domain and estimate the final super-resolved emissions through \eqref{eq:input adaptation}.
We exploit either $\DS$ or $\DSt$ datasets for training.
Since the $\simulated$ domain has higher native spatial resolution than the $\observed$, we train on different spatial resolutions,
addressing the $0.50^{\circ} \rightarrow 0.25^{\circ}$ case and its relative coarser version, from $1.00^{\circ}$ to $0.50^{\circ}$. 
Then, we test on $\DO$ datasets to pass from $1.00^{\circ}$ to $0.50^{\circ}$.

Fig.~\ref{fig:data adaptation} reports the system performance trend.
Training $\Tda$ with a small percentage of emission patches gives a slightly better generalization performance and captures the underlying statistics of the data more accurately. 
This is a counter-intuitive result since having more data for training a model generally improves its performance on unseen data. However, this effect consistently holds across multiple experiments, indicating the genuine nature of this phenomenon.


Training with higher spatial resolution $\simulated$ patches, i.e., from $0.50^{\circ}$ to $0.25^{\circ}$, leads to a significant improvement in the reconstruction of $\observed$ emission profiles. 
We believe 
higher spatial resolution in the training phase allows the network to learn high-frequency features that positively impact the reconstruction process. 

In addition, notice that exploiting $\Nets$, which uses emissions from $\DS$ dataset, has positive impact in reconstructing $\DO$ emissions compared to $\Netst$. This was expected, due to the huge number of patches used in training.
Nonetheless, considering also the training timings in Tab.~\ref{tab:no_da_results}, $\Netst$ results in slightly lower performance than $\Nets$ but leads to a substantial reduction in the computations.

Given these reasons, we select the model trained on the higher spatial resolution ($0.50^{\circ} \rightarrow 0.25^{\circ}$) and on $\DSt$ emissions as a baseline for subsequent \gls{da} experiments.



\begin{figure}[t]
\centering
\includegraphics[width=1\columnwidth]{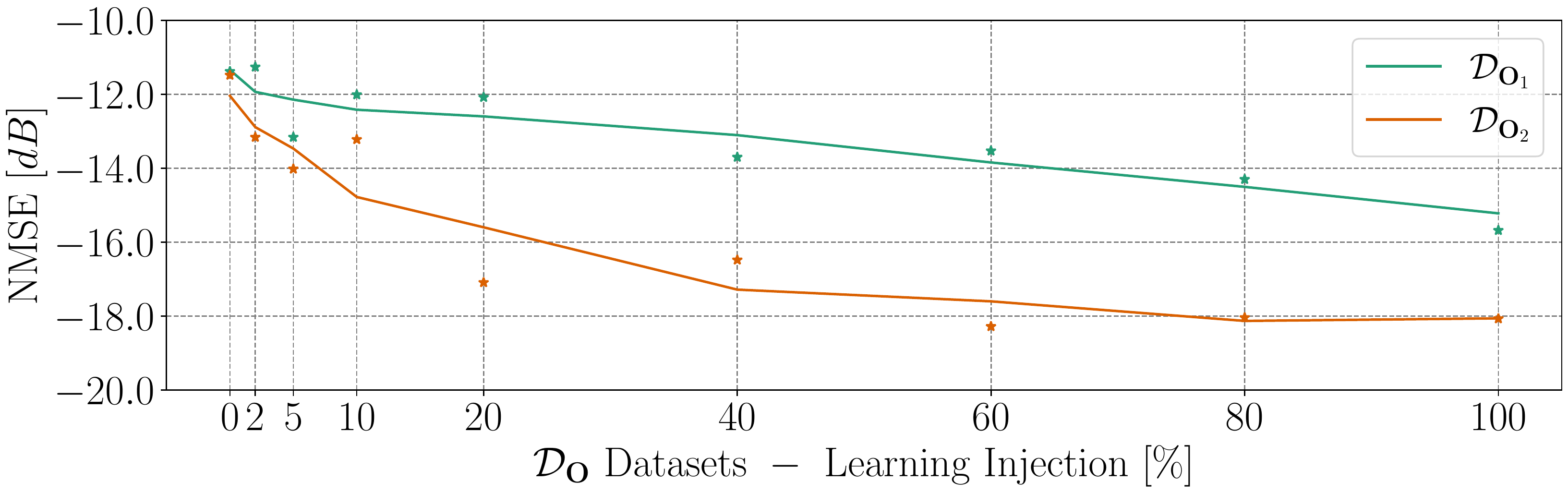}
\vspace{-17pt}
\caption{Average \gls{nmse} achieved by the \textit{Network Operator} adaptation strategy, in which we train the \gls{nn} operator by varying the amount of injected patches from the $\observed$ domain.}
\label{fig:network adaptation}
\vspace{-15pt}
\end{figure}

\noindent \textbf{Network Operator Adaptation. }
In this strategy, we systematically inject a portion of data belonging to $\DO$ datasets in our system's learning (training and validation) phase.
The remaining part of the dataset used in learning phase comprises emission patches belonging to the $\simulated$ domain, specifically to the $\DSt$ dataset.
For instance, the $0\%$ injection case corresponds to our preliminary investigations, where we use only data from the $\simulated$ domain for training the \gls{nn} operator.
Contrarily, the $100\%$ injection case is related to the last two rows of Table~\ref{tab:no_da_results}, which correspond to the perfect knowledge scenario. 

The \gls{nn} operator is not trained from scratch. We fine-tune the \gls{nn}, adapting the network weights pre-trained on $\simulated$ data ($\DSt$ dataset, from $0.50^{\circ}$ to $ 0.25^{\circ}$) to handle data from the target domain $\observed$, thus reducing the inherent domain-shift. For this reason, we refer to the new \gls{nn} operator as $\Netda$.
Regarding the data transformation applied to $\observed$ data, we select the best configuration found for the data transformation adaptation $\Tda$ (see Fig.~\ref{fig:data adaptation}). 
Final super-resolved emissions from $\observed$ domain are obtained using \eqref{eq:network adaptation}.

Fig.\ref{fig:network adaptation} reports the performances by varying the number of emission maps from $\observed$ domain used in the injection process. 
We can notice a relevant trend: increasing the percentage of data injection benefits the \gls{sr} quality for both the $\DO$ datasets. 
For $\DOuno$ emissions, just a $5\%$ injection gives an improvement of $1.8$ dB respect to the $0\%$ case. For $\DOdue$ emissions, a suitable trade-off between the amount of data and performance can be found when using $20\%$ injection, with an improvement of $5.6$ dB with respect to the $0\%$ case. Above the $60\%$ of data injection, results are comparable with the perfect knowledge scenario.
For the sake of clarity, we report at this link\footnote{\href{https://github.com/polimi-ispl/sr-bvoc}{https://github.com/polimi-ispl/sr-bvoc}} some additional results.

\vspace{-5pt}

\section{Conclusions}
\label{sec:conclusion}
\glsreset{da}
\glsreset{dl}
\glsreset{sr}

In this work, we tackled the problem of \gls{sr} of \gls{bvoc} emission maps derived from satellite observations.
Existing \gls{sr} methods are based on \gls{dl} and are effective on numerically simulated emission maps, thanks to the huge amount of data at disposal for training the models. 
This is not true when dealing with \gls{bvoc} emission maps of satellite-derived data which are often scarce, thus limiting the \gls{sr} performance. 

To solve this issue, we proposed to exploit the information coming from numerically simulated emission maps. However, some \gls{da} from the simulated to the observed data domain is necessary, since the two domains are characterized by extremely different spatial frequency behaviour and dynamic ranges.

Therefore, we proposed two \gls{da} strategies to solve the \gls{sr} of \gls{bvoc} emission maps of satellite-derived data.
First, we investigated a data transformation adaptation to adapt the different aggregation strategies and dynamics of simulated and observed data domains. Then, we explored a neural network adaptation, analyzing the influence of fine-tuning \gls{dl}-based models on real observations.  
Our experimental results shows that \gls{da} strategies are the key for achieving high quality reconstruction on emission maps derived from satellite observations, even in data scarcity conditions.
\vspace{-5pt}

\section{References}
\AtNextBibliography{\fontsize{8}{8}\selectfont}
\printbibliography[heading=none]

\end{document}